\documentclass[final,3p,times,12pt,fixfloat]{elsarticle}
\usepackage{epsfig}
\usepackage{slashed}
\usepackage{listings}
\usepackage{booktabs,multirow,tabularx}
\usepackage{amsmath}
\usepackage{float} % improved interface for floating objects
\usepackage{subfig} % supersedes the subfigure package
\usepackage{morefloats}
\usepackage{color}
\usepackage{multirow,bigdelim}
\usepackage{lineno}
\usepackage{tabularx}
\usepackage{subfig}
\usepackage{color}

\usepackage{footnote}
\biboptions{sort&compress}
\usepackage{lineno}
\journal{Nuclears Physics B}

\def\jpsi{J/\psi}
\def\psios{\psi'}

\def\etac{\eta_c}
\def\etacp{\eta_c'}

\def\etacts{\eta_c'}

\def\csetac{{\langle\mathcal{O}^{\etac}(\bigl.^1\! S_0^{[1]})\rangle}}

\newcommand{\ben}{\begin{enumerate}}
\newcommand{\een}{\end{enumerate}}
\newcommand{\bit}{\begin{itemize}}
\newcommand{\eit}{\end{itemize}}
\newcommand{\bc}{\begin{center}}
\newcommand{\ec}{\end{center}}
\newcommand{\bq}{\begin{equation}}
\newcommand{\eq}{\end{equation}}
\newcommand{\bqa}{\begin{eqnarray}}
\newcommand{\eqa}{\end{eqnarray}}

\newcommand{\ct}[1]{{Table~(\ref{#1})}}
\newcommand{\cf}[1]{{Fig.~\ref{#1}}}

\def\jpsi{{J/\psi}}

\def\chicj{{\chi_{cJ}}}
\def\chicz{{\chi_{c0}}}
\def\chico{{\chi_{c1}}}
\def\chict{{\chi_{c2}}}

\def\p0{{\bigl.^3\hspace{-1mm}P^{[8]}_0}}

\def\to{\rightarrow}

\def\MG5aMC{{\sc \small MadGraph5\_aMC@NLO}}

\def\jpsi{J/\psi}

\def\etac{\eta_c}
\def\etacp{\eta_c'}

\def\ppbar{p\bar{p}}
\def\BR{\mathcal{B}}
\newcommand{\Br}{{\cal B}}

\def\ie{{\it i.e.}}
\def\eg{{\it e.g.}}

\usepackage{ifpdf}
\ifpdf
\usepackage[pdftex]{hyperref}
\else
\usepackage[hypertex]{hyperref}
\fi

\hypersetup{
  pdftitle={Phenomenological NLO analysis of $\eta_c$ production at the LHC in the collider and fixed-target modes},%
  pdfauthor={Y. Feng, J. He, J.P. Lansberg, H.S. Shao, A. Usachov, H.F. Zhang},%
  pdfsubject={},%
  pdfkeywords={},%
  pdfstartview={},%
  bookmarksopen=true, breaklinks=true, debug=true, %
  colorlinks=true, linkcolor=red, citecolor=blue, urlcolor=blue
}

\newcommand{\glabcms}{\gamma^{\rm lab}_{\rm cms}}
\newcommand{\blabcms}{\beta^{\rm lab}_{\rm cms}}
\begin{document}

\begin{frontmatter}

\title{Phenomenological NLO analysis of $\eta_c$ production at the LHC in the collider and fixed-target modes}

\author[Chongqing]{Yu~Feng}
\author[LAL,UCAS]{Jibo~He}
\author[IPNO]{Jean-Philippe~Lansberg}
\author[LPTHE]{Hua-Sheng~Shao}
\author[LAL]{Andrii~Usachov}
\author[Guiyang]{Hong-Fei~Zhang}

\address[Chongqing]{Department of Physics, College of Basic Medical Sciences, Army Medical University, Chongqing 400038, China}
\address[LAL]{LAL, Universit\'e Paris-Saclay, Univ. Paris-Sud, CNRS/IN2P3, F-91898, Orsay Cedex, France}
\address[UCAS]{School of Physical Sciences, University of Chinese Academy of Sciences, 
19A Yuquan Road, Shijingshan district, Beijing 100049, P.R. China}
\address[IPNO]{IPNO, Universit\'e Paris-Saclay, Univ. Paris-Sud, CNRS/IN2P3, F-91406, Orsay Cedex, France}
\address[LPTHE]{LPTHE, UMR 7589, Sorbonne Universit\'es \& CNRS, F-75252, Paris Cedex 05, France}
\address[Guiyang]{College of Big Data Statistics, Guizhou University of Finance and Economics, Guiyang 550025, China}

\date{\today}

\begin{abstract}
In view of the good agreement between the LHCb prompt-$\eta_c$  data at $\sqrt{s}=$ 7 and 8 TeV and the NLO colour-singlet model predictions --\ie\ the leading $v^2$ NRQCD contribution--, we provide predictions in the LHCb acceptance for the forthcoming 13 TeV analysis bearing on data taken during the LHC Run2. We also provide predictions 
for $\sqrt{s}=115$ GeV for proton-hydrogen collisions in the fixed-target mode which could be studied during the LHC Run3. Our predictions are complemented by a full theoretical uncertainty analysis. In addition to cross section predictions, we elaborate on the uncertainties on the $p\bar p$ branching ratio --necessary for data-theory comparison-- and discuss other usable branching fractions for future studies.
\end{abstract}

\end{frontmatter}

\section{Introduction}

In 2014, LHCb released the first experimental study of prompt-$\etac$ hadroproduction at the LHC~\cite{Aaij:2014bga} at $\sqrt{s}=$ 7 and 8 TeV. 
It was found that the cross section measured by LHCb was compatible with a negligible
contribution of Colour-Octet (CO) transitions. More quantitatively, this observation combined with Heavy-Quark-Spin Symmetry (HQSS) yielded severe constraints on 
the corresponding CO transitions at work on $J/\psi$ production~\cite{Han:2014jya,Zhang:2014ybe,Butenschoen:2014dra,Likhoded:2014fta}. These are so stringent that only one fit~\cite{Han:2014jya} currently survives
these constraints at the expense of a slight tension with the CDF polarisation data~\cite{Abulencia:2007us}\footnote{The recent IHEP analysis where $\lambda_\phi$ and $\lambda_{\theta\phi}$ were computed for the first time
at NLO~\cite{Feng:2018ukp} also does not comply with the $\etac$ data.}. For reviews on quarkonium production, the reader is referred to Refs.~\cite{Brambilla:2010cs,Andronic:2015wma,Rapp:2008tf,Lansberg:2008gk,Lansberg:2006dh,Kramer:2001hh}

In this paper, we provide predictions for prompt-$\eta_c$ hadroproduction at $\sqrt{s}=$ 13 TeV 
to further test the compatibility between the Colour-Singlet (CS) contributions and the data and then in turn to refine the constraints on the Long-Distance Matrix Elements (LDMEs) associated with the dominant CO contributions. See~\cite{Lansberg:2017ozx} for a recent similar study
for the $\etacp$ case for which forthcoming data will also be invaluable. Since such constraints need to be extracted taking a proper account of both theoretical and experimental uncertainties, we also elaborate on our knowledge of the branching fractions of the decay channels which can be used by LHCb as well as on the scale and Parton-Distribution Functions (PDFs) uncertainties of the CS cross-section predictions.

The structure of the article is as follows. Section 2 is devoted to the discussion on the decay channels. Section 3 explains the theory framework we have used to provide CS NLO predictions and gathers our predictions both for the collider kinematics and for the fixed-target kinematics. Section 4 gathers our conclusion.

\section{Discussion on the decay channels}

The decays of non-$1^{--}$ charmonium states to the experimentally  clean di-muon channel are strongly suppressed and hence these states can only be reconstructed  using decays to hadrons or their radiative transitions to underlying charmonium states. In this section we discuss possible decay channels to study $\etac$, $h_c$ and $\etacts$, which cannot be accessed using their decays to $\mu^+\mu^-$ or $\jpsi \gamma$. The known branching fractions~\cite{PDG2017} of the decays discussed below are summarised in Tab.~\ref{tab:Brs}. Many of these  branching fractions can be measured more precisely at Belle, Belle II, BES III, or the super tau-charm experiments.

\begin{savenotes}
\begin{table}[h!]
\newsavebox{\tablebox}
\begin{center}
\begin{lrbox}{\tablebox}
\begin{tabular}{c||c|c|c|c|c|c|c|c|c}
                  
& \multicolumn{8}{ c |}{$\BR\times 10^3$}\\
&$\ppbar$
&$\phi\phi$ 
&$\phi K^+K^-$ 
&$\phi\pi^+\pi^-$
&$\Lambda\overline{\Lambda}$ 
&$\Xi^+{\Xi}^-$ 
&$\Lambda(1520)\overline{\Lambda}(1520)$
&$\etac\gamma$
&$\ppbar\pi^+\pi^-$
\\
\hline \hline 
$\etac$ 
&$1.52\pm0.16$ 			%ppbar
&$1.79\pm0.20$ 			% phiphi
&$2.9\pm1.4$ 			% phiKK
&unknown 				% phipipi
&$1.09\pm0.24$ 			%LambdaLambda
&$0.90\pm0.26$ 			% XiXi
& - 					% LstLst
& - 					% etac gamma
&$5.3\pm1.8$			% ppbarpipi
\\				

$\jpsi$ 
&$2.12\pm0.03$ 			%ppbar
&forbidden 				% phiphi
&$0.83\pm0.12$ 	% phiKK
&$0.87\pm0.09$ 			% phipipi
&$1.89\pm0.08$			%LambdaLambda
&$0.97\pm0.08$			% XiXi
&unknown 				% LstLst
&$17\pm4$ 				% etac gamma
&$6.0\pm0.5$ 			% ppbarpipi
\\

$\chicz$ 
&$0.22\pm0.01$			%ppbar
&$0.80\pm0.07$			% phiphi
&$0.97\pm0.25$			% phiKK
&unknown				% phipipi
&$0.33\pm0.02$			%LambdaLambda
&$0.48\pm0.07$			% XiXi
&$0.31\pm0.12$ 			% LstLst
&forbidden				% etac gamma	
&$2.1\pm0.7$			% ppbarpipi
\\

$h_c$ 
&$<0.15$				%ppbar 
&forbidden          	% phiphi
&unknown				% phiKK
&unknown 				% phipipi
&unknown				%LambdaLambda
&unknown				% XiXi
&unknown 				% LstLst
&$510\pm60$ 			% etac gamma
&unknown	  			% ppbarpipi
\\

$\chico$ 
&$0.076\pm0.003$		%ppbar
&$0.42\pm0.05$			% phiphi
&$0.41\pm0.15$			% phiKK
&unknown				% phipipi
&$0.11\pm0.01$		%LambdaLambda
&$0.08\pm0.02$		% XiXi
&$<0.09$  				% LstLst
&forbidden				% etac gamma
&$0.50\pm0.19$  		% ppbarpipi
\\

$\chict$ 
&$0.073\pm0.003$		%ppbar
&$1.06\pm0.09$			% phiphi
&$1.42\pm0.29$			% phiKK
&unknown				% phipipi
&$0.18\pm0.02$		%LambdaLambda
&$0.14\pm0.03$		% XiXi
&$0.46\pm0.15$  		% LstLst
&forbidden				% etac gamma
&$1.32\pm0.34$			% ppbarpipi
\\

$\etacts$				
&$0.07$\footnote{Indirect determination} 
&unknown          		% phiphi
&unknown				% phiKK
&unknown 				% phipipi
&unknown				%LambdaLambda
&unknown				% XiXi
&unknown 				% LstLst
&forbidden 				% etac gamma
&unknown    			% ppbarpipi
\\

$\psios$ 
&$0.29\pm0.01$		%ppbar
&forbidden				% phiphi
&$0.07\pm0.02$		% phiKK
&$0.12\pm0.03$		% phipipi
&$0.38\pm0.01$		%LambdaLambda
&$0.29\pm0.01$		% XiXi 
&unknown 				% LstLst
&$3.4\pm0.5$			% etac gamma
&$0.60\pm0.04$			% ppbarpipi
\\
\end{tabular}
\end{lrbox}
\resizebox{1.0\textwidth}{!}{\usebox{\tablebox}}
\end{center} 
\vskip-0.6cm
\caption{The branching fractions ($\times 10^3$) of charmonium decays to hadrons and radiative decays to $\etac \gamma$. }
\label{tab:Brs}
\end{table} 
\end{savenotes}

The $\ppbar$ decays of charmonia have been investigated as a possible channel to measure charmonium production at the LHC~\cite{Barsuk:2012ic}.
The measurement of the $\etac$ production at the LHCb experiment has been performed using the $\etac \to \ppbar$ decay~\cite{Aaij:2014bga}, which demonstrated that the $\ppbar$ final state is 
powerful to reconstruct the $\etac$ state and measure the $\etac$ production rate relative to that of the $\jpsi$, even though the $\etac$ hadroproduction rate is measured only for $\etac$ with transverse momenta ($P_T$) larger than 6.5 GeV due to the available trigger bandwidth. Also, this decay is used to study exotic candidates decaying to $\etac \pi^-$~\cite{Aaij:2018bla}. The branching fraction of the $\etac \to \ppbar$ is known to about 10\% precision~\cite{PDG2017}. The studies of the $\etac$ would benefit from a more precise measurement of $\BR(\etac \to \ppbar)$ or $\BR(\etac \to \ppbar)/\BR(\jpsi \to \ppbar)$. Branching fractions of $\chicj \to \ppbar$ decays and $\psios \to \ppbar$ have been measured to about 3-5\% precision. Recently, LHCb has observed the decay $\etacts \to \ppbar$ using a data sample of exclusive $B^+ \to \ppbar K^+$ decays~\cite{Aaij:2016kxn}. Together with the measurement of $\BR(B^+ \to \etacts K^+)$ by Belle~\cite{Kato:2017gfv}, the branching fraction of $\etacts \to \ppbar$ is indirectly determined to be about $0.7\times10^{-4}$. Therefore, the decay $\etacts \to \ppbar$ is promising for the $\etacts$ hadroproduction measurement.

The other promising final state to study prompt production of charmonium is $\phi\phi$. The $1^{-}$ charmonium states are forbidden to decay to $\phi\phi$. LHCb measured the $\chi_{c0,1,2}$ and $\etacts$ production in inclusive $b$-hadron decays using the $\phi\phi$ final state with the first evidence of the $\etacts \to \phi\phi$ decay~\cite{Aaij:2017tzn}.  In the latter analysis, a possible problem was highlighted, namely  the PDG fit value of $\BR(\etac \to \phi\phi)$ differs from the PDG average value~\cite{PDG2017} by a factor close to 2. In addition, the ratio of the branching fractions $\BR(\etac \to \phi\phi)/\BR(\etac \to \ppbar)$ was measured. More measurements are needed to establish a robust value of the $\BR(\etac \to \phi\phi)$. Also, due to the evidence of the $\etacts \to \phi\phi$, this channel is promising to measure the hadroproduction of the $\etacts$. Similarly, the $\phi K^+ K^-$ and the $\phi \pi^+ \pi^-$ final states could be used.

The branching fractions of charmonium decays to long-lived baryons such as $\Lambda\bar{\Lambda}$ and $\Xi^+\Xi^-$ are measured for most charmonium states. The reconstruction of these decay channels is challenging for LHCb due to the large lifetimes of these baryons such that they escape the Vertex Locator (VELO), which cause a small reconstruction and trigger efficiency. Decays involving short-lived baryons are reconstructed by LHCb with better efficiency. 

The decays $\chi_{c0,2} \to \Lambda(1520)\bar{\Lambda}(1520)$ have been observed by the BES III collaboration~\cite{Ablikim:2011uf} while the $\jpsi \to \Lambda(1520)\bar{\Lambda}(1520)$ decay is not observed so far. This channel becomes another candidate to measure hadroproduction of charmonium states~\cite{Jacques}.

The least studied charmonium state is the $h_c$ meson, and not many of $h_c$ decays have been observed so far. The $h_c$ meson is expected to decay to $\ppbar$, however, the upper limit on the $\BR(h_c \to \ppbar)$ reported by the BES III collaboration~\cite{Ablikim:2013hdv} is more than one order of magnitude smaller than the theoretical prediction~\cite{Barsuk:2012ic}.  Also, the $h_c$ can be measured using its radiative transition $h_c \to \etac \gamma$ with branching fraction about 50\%, which requires reconstruction of the $\etac$ state. Recently, LHCb observed the very clean decays $\chi_{c1,2} \to \jpsi \mu^+ \mu^-$, and precisely measured  the $\chict$ mass and its natural width~\cite{Aaij:2017vck}. The $h_c \to \etac \mu^+ \mu^-$ decay can be searched similarly. Also, the BES III has observed the $h_c \to \ppbar \pi^+\pi^-$ decay and measured its branching fraction~\cite{Ablikim:2018ewr} to be $(2.89\pm0.32\pm0.55)\times10^{-3}$, which is promising for searches by LHCb.

\section{Framework and results}

\subsection{Framework}

The present NLO analysis was performed  thanks to the FDC framework~\cite{Wang:2004du,Gong:2014qya}\footnote{The FDC (standing for Feynman Diagram Calculation) package  have been developed to automated HEP computations. It is based on the LISP symbolic programming language in order to produce FORTRAN
codes. The Lagrangian are formed by the code, following the user requirement, from which are derived the corresponding Feynman rules. The package generates all possible Feynman diagrams contributing to a
given process up to one loop in a given model. It can in particular deal quarkonium production within NRQCD. The amplitude 
of the process are analytically manipulated to generate FORTRAN codes of the squared amplitudes up to one loop. 
Numerical results for the (differential) cross sections are then computed by performing  the phase-space integrals
using the phase-space slicing method. We refer to~\cite{Gong:2014qya} for explanations relevant to quarkonium-production applications.} which generates the Born, real-emission
and virtual contributions, ensures the finiteness of their sum, performs the partonic-phase-space integration and that
over the PDFs. As announced, we performed a full study of the scale uncertainty by varying both $\mu_R$ and $\mu_F$ about the default
value $\mu_0=\sqrt{m_{\eta_c}^2+P_T^2}$ as $(\mu_R,\mu_F)=\mu_0 \times (1,1;0.5,0.5;2,2;0.5,1;1,0.5;1,2;2,1)$.

As for the CS LDME, we have taken $\csetac=0.39$~GeV$^3$ which corresponds to $|R(0)|^2=0.81$~GeV$^3$ for
the radial wave function at the origin. In order to study the impact of the PDF uncertainties at NLO, we have used the CT14 set~\cite{Dulat:2015mca} which is included in LHAPDF5~\cite{Whalley:2005nh}. 
The corresponding uncertainties follow from the 57 eigensets of CT14.

\subsection{Results for the collider mode at $\sqrt{s}=13$ {\rm TeV}} 

Our predictions at $\sqrt{s}=13$ {\rm TeV} follow from the expected kinematical range of the forthcoming LHCb study performed on 
data taken during the Run2 in 2015-2016. They correspond to 2~fb$^{-1}$ of data at $\sqrt{s}=13$ {\rm TeV}. We have therefore considered
the same rapidity acceptance as that used for the first LHCb analysis~\cite{Aaij:2014bga}, namely $2< y_{\rm cms} < 4.5$ without any additional
fiducial cuts on the decay product of the $\eta_c$. 

\cf{fig:dsigspt13TeV} displays our predictions for the $P_T$-differential prompt-$\eta_c$ cross section at NLO accuracy along with their associated scale and PDF uncertainties. It is clear that the latter are negligible in this energy range as compared to those from the scales. \cf{fig:Kfactor13TeV} shows the ratio of the NLO/LO cross sections with the scale uncertainties only and points at a $K$ factor slightly increasing with $P_T$. This is the expected behaviour with leading $P_T$ channels opening up at $\alpha_s^4$. It also shows that the scale uncertainty is as large as 50\%.

\begin{figure}[hbt!]
\begin{center}
\subfloat[]{\includegraphics[width=0.5\textwidth,draft=false]{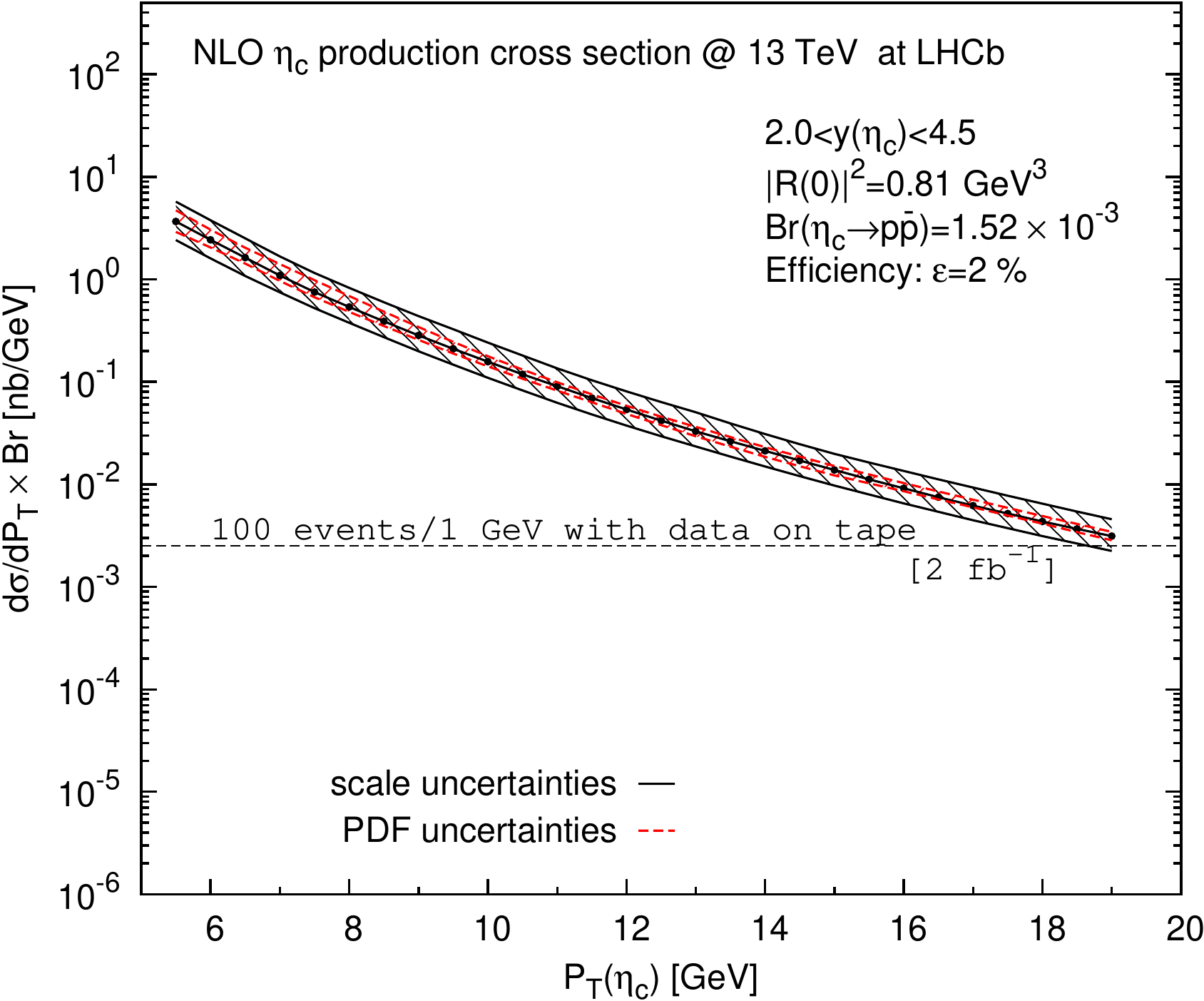}\label{fig:dsigspt13TeV}}
\subfloat[]{\includegraphics[width=0.475\textwidth,draft=false]{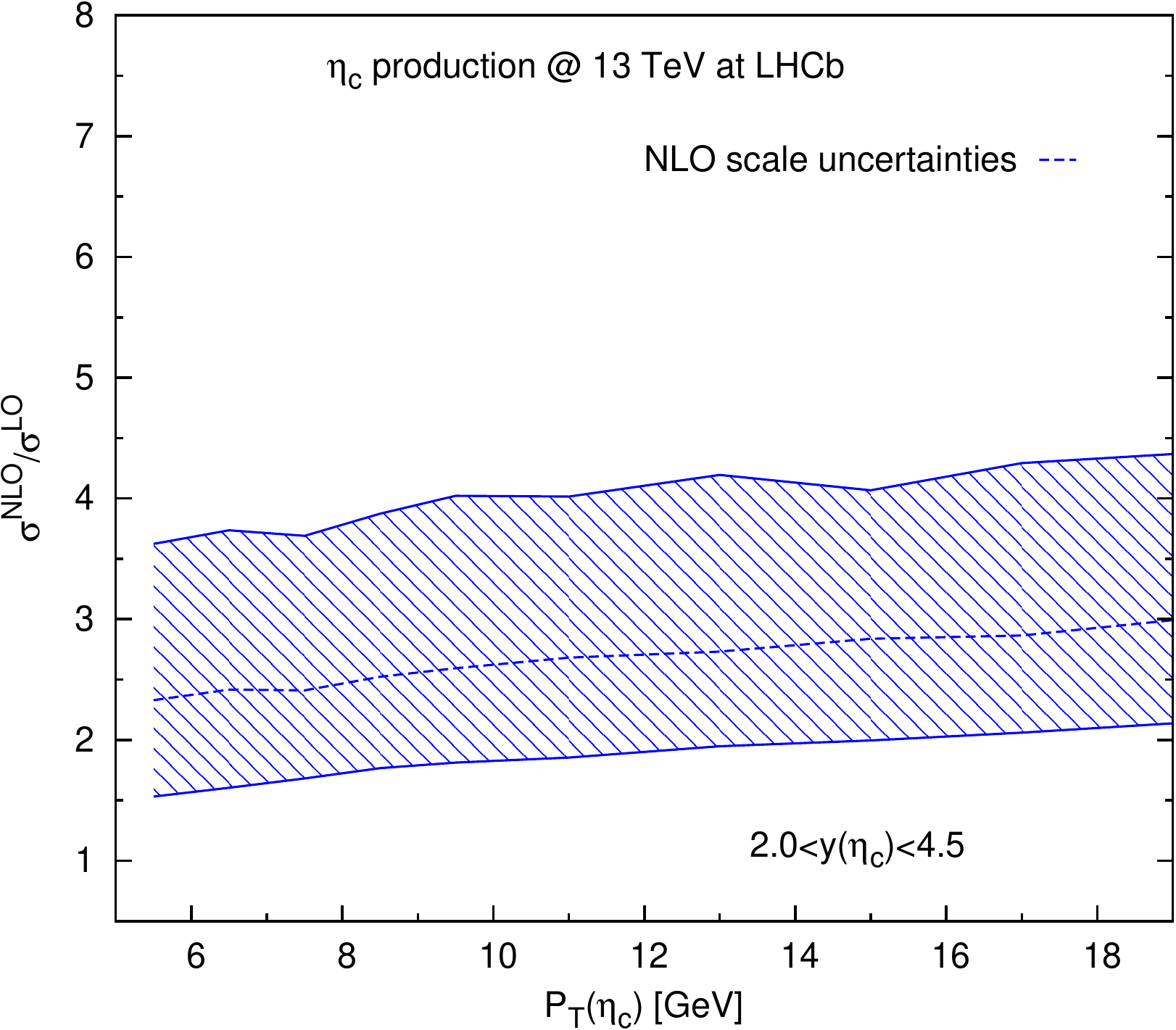}\label{fig:Kfactor13TeV}}
\caption{\label{fig:13TeV} (a) NLO $P_T$ differentical cross section in the LHCb acceptance at 13 TeV. The black (red) hatched band denotes the scales (PDF) uncertainties; (b) NLO/LO cross-section ratio as a function of $P_T$ where  only the scale uncertainty on the NLO cross section is shown.}
\end{center}
\end{figure}

Assuming a recorded luminosity of 2~fb$^{-1}$, $\Br(\eta_c \to p \bar p)=1.52 \times 10^{-3}$ and an efficiency on the order of 2 \%, the 100-count limit per GeV correspond to 2~pb and located around $P_T\simeq 20$~GeV. Without any surprise, the increase in the energy should allow LHCb to push their measurements at 13 TeV to slightly larger $P_T$ compared to 7 and 8 TeV. Limitation may come from the range where the $J/\psi\to p \bar p$ yield is measured as well as from systematical uncertainties. In view of the other branching fractions on~\ct{tab:Brs}, let us add that other decays are also within the reach of LHCb measurements.

\subsection{Results for the fixed-target mode at $\sqrt{s}=115$ {\rm GeV}}

The use of the proton LHC beam in the fixed-target mode has lately be the object of intense
investigation both in terms of feasibility and in terms of physics reach, see \eg~\cite{Hadjidakis:2018ifr,Lansberg:2018fsy,Karpenko:2018xam,Begun:2018efg,Goncalves:2018htp,Massacrier:2017lib,Massacrier:2015qba,Trzeciak:2017csa,Kikola:2017hnp,Anselmino:2015eoa,Lansberg:2015lva,Arleo:2015lja,Zhou:2015wea,Lansberg:2015kha,Goncalves:2015hra,Kanazawa:2015fia,Vogt:2015dva,Ceccopieri:2015rha,Chen:2014hqa,Rakotozafindrabe:2013au,Rakotozafindrabe:2012ei,Lansberg:2012kf,Brodsky:2012vg}. 
In particular, a wide variety of measurements in different possible implementations were discussed in~\cite{Hadjidakis:2018ifr}. We will limit ourselves here to a few statements on the kinematics. First, 7 TeV protons impinging fixed targets release  a center-of-mass system (cms)
energy close to 115~GeV ($\sqrt{s}=\sqrt{2E_p m_N}$). Second,  the boost between the cms
and the laboratory is $\glabcms=\sqrt{s}/(2m_p)\simeq 60$ yielding a rapidity
shift as large as $\tanh^{-1} \blabcms\simeq 4.8$. As such, the nominal acceptance of the LHCb detector in the cms approximates to $-2.8 < y_{\rm cms} < -0.3$. Physics wise, the LHCb detector probes backward physics in the fixed-target mode.

Nowadays, the first analysed fixed-target data based on the SMOG LHCb system --initially designed to improve the luminosity determination in LHCb, now used as a low-density-unpolarised-gas target-- are coming in~\cite{Aaij:2018ogq,Aaij:2018svt} and confirm that the particle multiplicity in the LHCb detector is such that its performance in the fixed-target mode remains intact. One can thus consider that similar decay channels of the $\eta_c$ as those discussed for the collider mode could be studied if sufficient luminosities can be achieved. 

Until now, the LHCb-SMOG statistical samples for $J/\psi$ taken with different
noble gases (He, Ar, Ne) remain too small --on the order of hundreds-- to expect any $\eta_c$ counts. The situation could significantly get better in the future with proposed SMOG2 system~\cite{Redaelli:2018oqa,Dainese:2019xrz} with achievable yearly luminosities on the order of 10 pb$^{-1}$ during the LHC Run3. It is however crucial to further constrain NRQCD LDMEs --as we propose here-- to have a H target available as opposed as to nuclear --noble gas-- targets. For the LHC Run4, yearly luminosities as high as few fb$^{-1}$ will be within experimental reach as discussed in~\cite{Hadjidakis:2018ifr}.

\cf{fig:dsigspt13TeV} displays the $P_T$-differential $\eta_c$ cross section at $\sqrt{s}=115$ {\rm GeV} in the expected acceptance of LHCb in the fixed-target mode. As above, we separated out the uncertainties from the scale variations ($\mu_F$ and $\mu_R$)  and from the PDFs which are a little larger here since one probes slightly larger $x$ values. As a matter of fact, dedicated rapidity-differential measurements at very negative $y_{\rm cms}$ could provide specific constraints on the gluon PDFs~\cite{Hadjidakis:2018ifr,Lansberg:2012kf,Brodsky:2012vg}. \cf{fig:Kfactor115GeV} shows the ratio of the NLO/LO cross sections with the scale uncertainties only and points at a $K$ factor slightly increasing with $P_T$. This is the expected behaviour with leading $P_T$ channels opening up at $\alpha_s^4$. It also shows that the scale uncertainty is as large as 5 at low energies.

\begin{figure}[hbt!]
\begin{center}
\subfloat{\includegraphics[width=0.5\textwidth,draft=false]{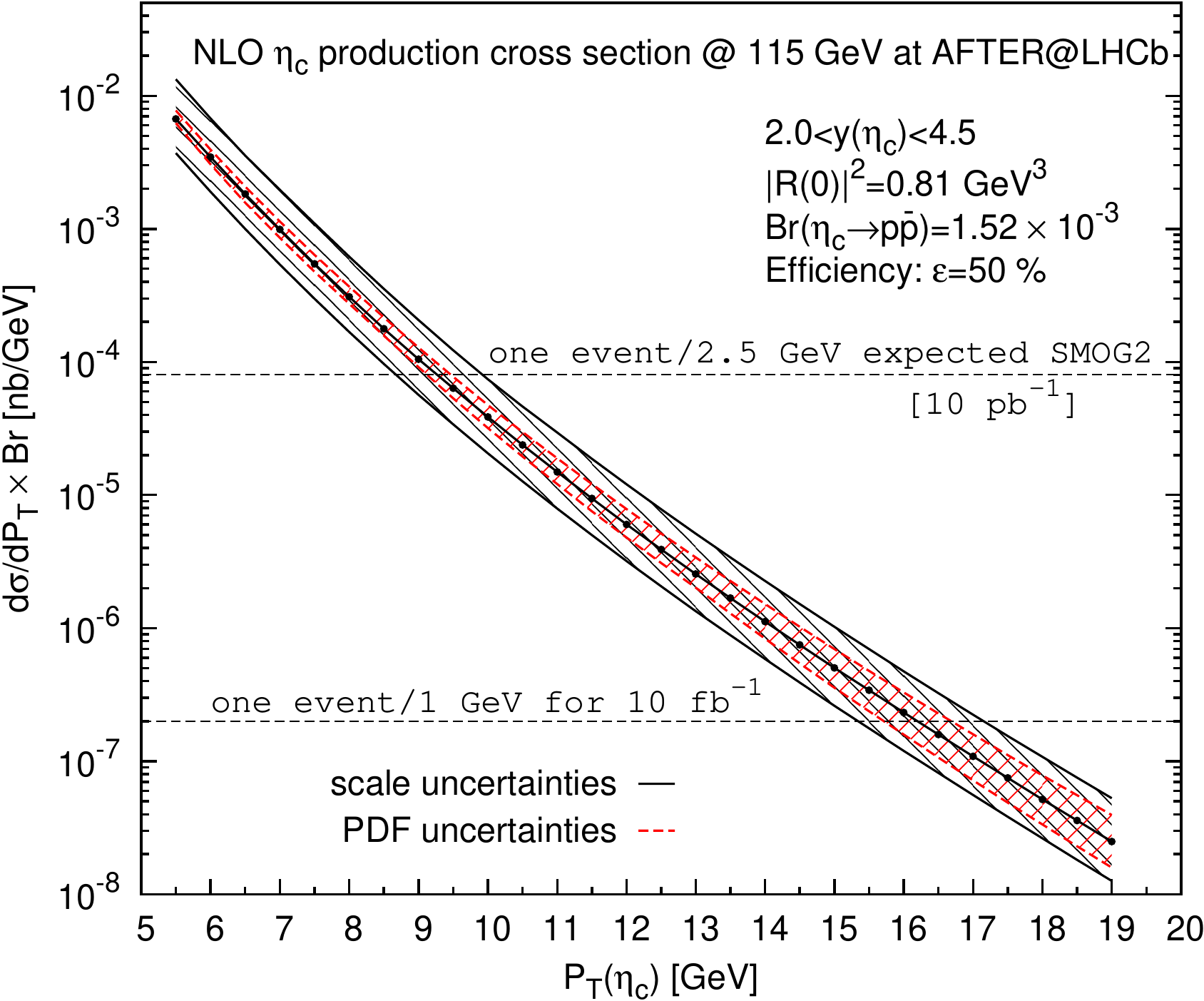}\label{fig:dsigspt115GeV} }
\subfloat{\includegraphics[width=0.475\textwidth,draft=false]{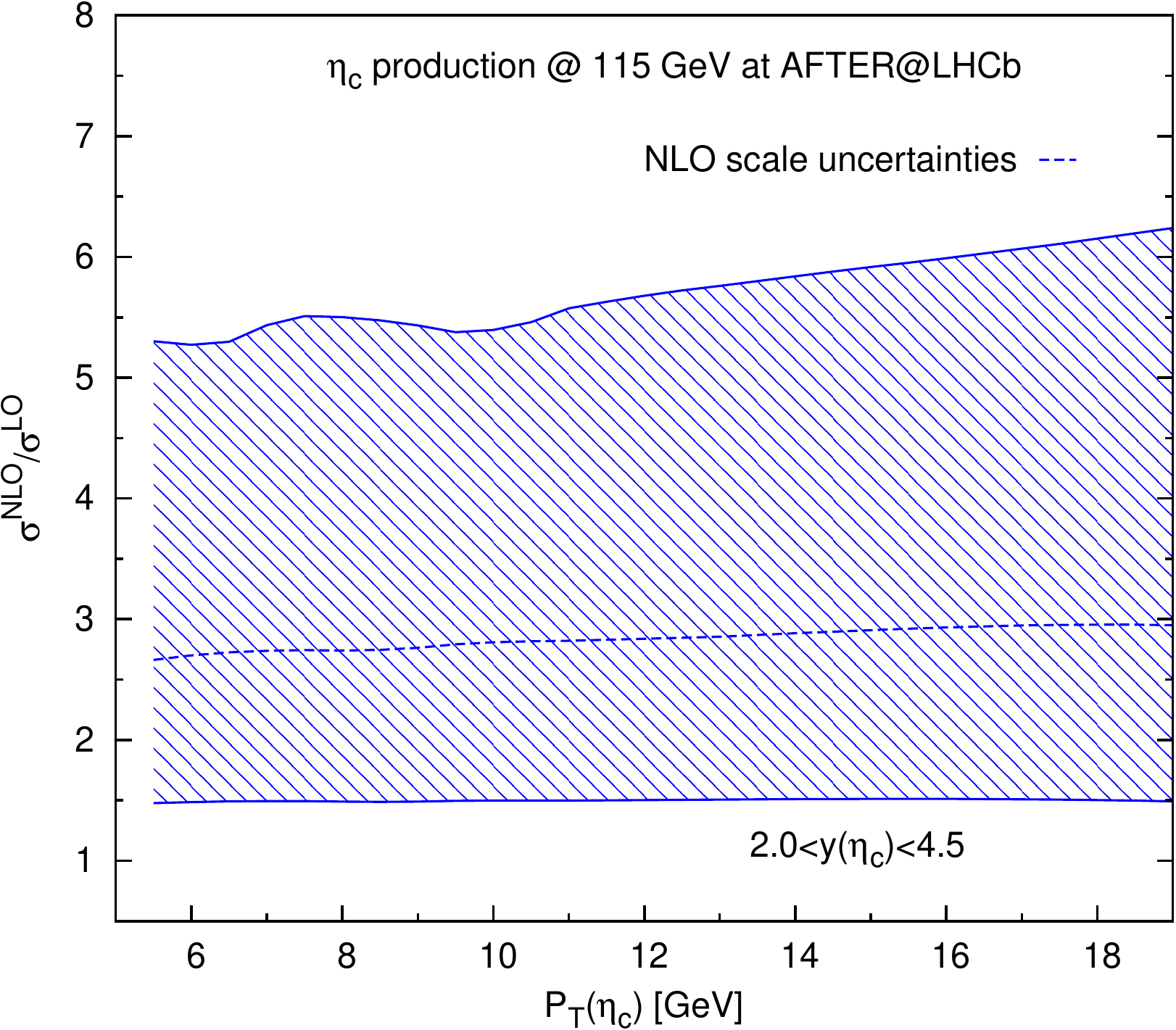}\label{fig:Kfactor115GeV} }
\caption{\label{fig:115GeV} (a) NLO $P_T$ differentical cross section in the LHCb acceptance at 115 GeV in the fixed-target mode. The black (red) hatched band denotes the scales (PDF) uncertainties; (b) NLO/LO cross-section ratio as a function of $P_T$ where  only the scale uncertainty on the NLO cross section is shown.}
\end{center}
\end{figure}
 	
Assuming an integrated luminosity of 10 pb$^{-1}$, $\Br(\eta_c \to p \bar p)=1.52 \times 10^{-3}$ and an efficiency on 50 \%, the one-count limit per 2.5 GeV for $d\sigma/dP_T$ is on the order 0.08 pb, which corresponds according to our results to a $P_T$ upper limit of $8.5 \div 10$~GeV. It precisely happens to be the range accessed at 7 and 8 TeV. With 10 fb$^{-1}$, the reach would simply be equivalent to that of the collider mode. We further note that, thanks to the reduced multiplicities in fixed-target mode, lower $P_T$'s should be accessible. This would allow one to measure the gluon Transverse-Momentum-Dependent functions (TMDs) along the lines of~\cite{Boer:2012bt,Ma:2012hh,Signori:2016jwo}.

\section{Conclusions and outlook\label{sec:con}}
We have computed the prompt $\eta_c$-production cross section at one loop accuracy in QCD and in the CSM (LO in $v^2$ of NRQCD) for the LHCb kinematics in the collider
mode at $\sqrt{s}=13$~TeV and in the fixed-target mode at $\sqrt{s}=115$ GeV. In addition, we have provided an up-to-date
discussion of the possible decay channels to be used for such studies and performed an original analysis of the theoretical
analysis including that from the factorisation and renormalisation scales and from the PDFs.

In addition, let us stress that the understanding and the measurements of $\eta_c$ production go well beyond the
determination of NRQCD LDMEs. Its production in proton-nucleus collisions (see \cite{Lansberg:2016deg} for predictions 
of the corresponding nuclear modification factors at LHC energies) can provide complementary means to probe the distribution of gluons inside nuclei
along the lines of~\cite{Kusina:2017gkz,Kusina:2018pbp}. In proton-deuteron collisions at extreme $x_F$, it can also give us some handle 
on the gluon distribution in the deuteron at very large $x$~\cite{Brodsky:2018zdh}.

\section*{Acknowledgements} 
We thank S. Barsuk for useful discussions. 
The work of YF, JH, JPL, HSS, HFZ is supported in part by CNRS via the LIA FCPPL. JPL is supported in part by the TMD@NLO IN2P3 project. HSS is supported in part by the LABEX ILP (ANR-11-IDEX-0004-02, ANR-10-LABX-63).

\bibliographystyle{utphys}

\bibliography{etac-LHCb-2018-JPL-191218}

\end{document}